# Construction of Learning Path Using Ant Colony Optimization from a Frequent Pattern Graph

Souvik Sengupta[1], Sandipan Sahu[2], Ranjan Dasgupta[3]

[1,2]Bengal Institute of Technology, Kolkata, India
[3]National Institute of Technical Teachers Training and Research, Kolkata, India

## 1. Abstract :

*In an e-Learning system a learner may come across multiple unknown terms, which are generally hyperlinked, while reading a text definition or theory on any topic. It becomes even harder when one tries to understand those unknown terms through further such links and they again find some new terms that have new links. As a consequence they get confused where to initiate from and what are the prerequisites. So it is very obvious for the learner to make a choice of what should be learnt before what. In this paper we have taken the data mining based frequent pattern graph model to define the association and sequencing between the words and then adopted the Ant Colony Optimization, an artificial intelligence approach, to derive a searching technique to obtain an efficient and optimized learning path to reach to a unknown term.*

## 2. Key words : e-Learning, Learning Path, Data Mining, Frequent Pattern , AI, ACO

## 3. Introduction :

The self regulated learning for definition of a term is where the learners hold incremental beliefs about intelligence within their control. In the hypertext document, links have been established in such a way that the user can explore, browse and search for not only a particular term but can also get information regarding relevant and associated issues [3]. The problem with self-regulated definition learning is that learner often gets confused where to start from because the term definition itself contains multiple keywords which are known or unknown. So prerequisites required to understand a particular term are most important consideration here. That is, ultimately the learner always tries to find out a learning path which is a sequence of words like $a_1\ a_2\ \dots\ a_n$. This sequence from source to destination may include several unknown terms which indicate the need of prerequisite knowledge required to understand the target term. In this paper first we have created a graph to show the interconnection between the terms. Then we have adopted a hybrid model in the sense that we take data from direct mode classroom question answer session between students and teachers and then train the graph data based on our experience. We have used the association rule of data mining to create the frequent pattern among the links of the graph and then derived the learning path using ACO technique.

## 4. Related works:

Many research works have been made to establish an individual learning path for the learners in which the application of Artificial Intelligence, Fuzzy-neural and Data Mining techniques are most commonly used. An AI based tutoring system [5] has been proposed to identify, monitor and adapt the student's learning path, according to the students actual knowledge, learning habits and preferred learning style, whereas personalized curriculum sequencing is another important research issue for Web-based learning systems because no fixed learning paths will be appropriate for all learners. Therefore, many researchers focused on developing e-learning systems with personalized learning mechanisms to assist online Web-based learning and adaptively provide learning paths in order to promote the learning performance of individual learners [6].

Experimental results indicated that applying a genetic-based personalized e-learning system for Web-based learning is superior to the freely browsing learning mode because of high quality and concise learning path for individual learners. In an another approach working on design of Learners' Quanta (LQ) based efficient and adaptive learning system, it has been observed that a generic platform is required to design and develop an LMS with the adaptive algorithm already proposed in [8], [9] and [10].

**A graph based frequent pattern model:**

Let in an e-Learning context a term 'A' is defined with help of terms 'B','C','D'. These terms may not be not in any order that means the sequence of these terms may come differently in a different context. So the order of the terms is not an important consideration here we have only concentrated on their presence as a prerequisite to describe a keyword. So we create a sequence like this: B-C-D-A and join this branch with root.

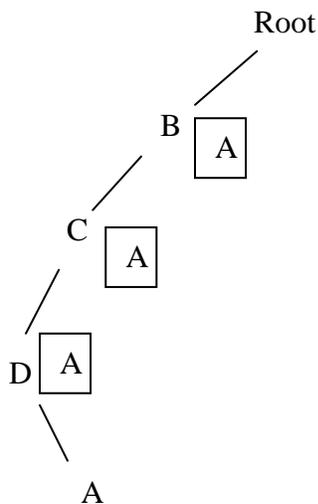

**fig 5.1 : branch BCDA**

We introduce an additional data structure called a datalist associated with each node that contains information about the term for which it has been used. The datalist keeps on updating as new branches are introduced because same term can be reused for describing multiple keywords. For instance if another term definition 'G' involves 'E','F', 'C' then we create another branch E-F-C-G and append it to the graph like fig 5.2 . now the node 'C' has 'A' and 'G' in its data list which indicates that 'C' is used to describe both these terms.

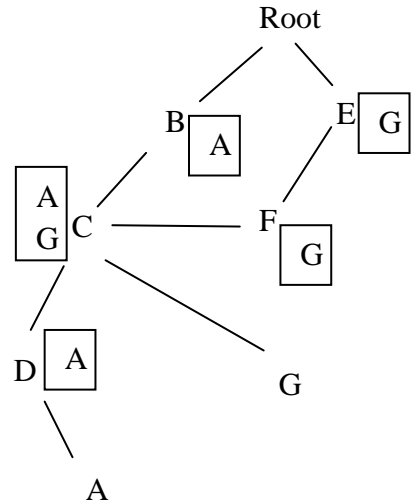

**fig 5.2: Appending EFCG**

This way we go on expanding the graph with new terms and link them with the existing terms.

### 5. Data Mining :

A popular and effective way of discovering new knowledge from large and complex data sets is data mining [1]. The association rule and the frequent pattern tree of data mining are the major efficient techniques in e-Learning data [4]. We will use association rule to reveal the dependencies of a term on the other term, the FP graph shows the potential usage of the other terms to learn any particular 'term' by the learner.

### 6.1 Creating frequent pattern and association :

We have taken the following examples of definition from a well known science dictionary website.

**DNA**: Deoxyribonucleic acid or DNA is a double-stranded **nucleic acid** that contains the genetic information for **cell** growth, division, and function.

**Nucleic Acid:** Nucleic acids direct the course of **protein** synthesis, thereby regulating all **cell** activities. The two main types, **DNA** and **RNA**, are composed of similar materials but differ in structure and function. Both are long chains of repeating **nucleotides**.

**Nucleotides:** Nucleotides are molecules that, when joined together, make up the structural units of **RNA** and **DNA**. A nucleotide consists of a sugar molecule attached to a phosphate group and a nitrogen-containing base. The bases are adenine, cytosine, guanine, and thymine or uracil (for RNA).

**RNA:** Ribonucleic acid or RNA a **nucleic acid** that is generally single stranded and plays a role in transferring information from **DNA** to **protein**-forming system of the **cell**.

**Protein:** Proteins are natural polymer molecules consisting of **amino acid** units. The number of amino acids in proteins may range from two to several thousand.

**Amino acid:** A **molecule** consisting of the basic amino group (NH2), the acidic carboxylic group (COOH), a hydrogen atom (H), and a side-chain(R) that varies between different amino acids attached to the carbon **atom**.

**Cell:** The structural, functional and biological unit of all organisms. It is the smallest unit of life that is classified as a living thing.

**Eukaryotic:** A **cell** that contains membrane-bound compartments in which specific **metabolic** activities take place. Most important among these compartments is the **nucleus**.

**Nucleus:** The nucleus is a membrane-enclosed **organelle** found in **eukaryotic cells**. It contains most of the cell's **genetic material**, organized as multiple long linear **DNA** molecules in complex with a large variety of **proteins**.

**Organelle:** An organelle is a specialized subunit within a **cell** that has a specific function.

**Genetic material:** The genetic material of a cell or an organism refers to those materials found in the nucleus, mitochondria and cytoplasm, which play a fundamental role in determining the structure and nature of cell substances.

**Mitochondria:** Mitochondria are the powerhouses of the human **cell**; mitochondrion is a membrane-enclosed **organelle** found in most **eukaryotic** cells. They generate most of the cell's supply of adenosine triphosphate (**ATP**), used as a source of chemical energy.

**Cytoplasm:** The cytoplasm is a thick liquid residing between the cell membrane holding all **organelles**, except for the **nucleus**. All the contents of the **cells** of **eukaryotic** organisms (which lack a **cell nucleus**) are contained within the cytoplasm.

**Ribosome:** A ribosome is the component of a biological **cell** that creates proteins from all **amino acids** and **RNA** representing the **protein**.

Next we create the transaction out of these terms by fetching out the keywords from the definitions.

| | |
|---|---|
| DNA: | *Cell, Nucleic acid.* |
| Protein: | *Amino acid.* |
| Eukaryotic: | *Cell, Metabolism, Nucleus.* |
| Mitochondria: | *Cell, Eukaryotic, organelle*. |
| Nucleic acid: | *Cell, DNA, Nucleotide, Protein, RNA.* |
| RNA: | *Cell, DNA, Nucleic acid, Protein.* |
| Nucleotide: | *DNA, RNA* |
| Amino acid: | *atom, Molecule* |
| Nucleus: | *DNA, eukaryotic, genetic material, organelle, proteins* |
| Organelle: | *Cell* |
| Genetic material: | *Cell, Cytoplasm, Mitochondria, Nucleus, Organism.* |
| Cytoplasm: | *Cell, Eukaryotic, Nucleus, Organelle,* |
| Ribosome: | *Amino acid, Cell, Protein, RNA.* |

### 6.2 Construction of the graph

We will now simulate the above sequences to form the branches of the graph. For instance the sequence of DNA will be Root--Cell--Nucleic Acid--DNA. Now as the branches are joined by the common terms, ultimately we form a frequent pattern graph by showing the frequency on individual edges. Fig 6.1 describes this graph, the frequencies are not shown to make the graph look simple.

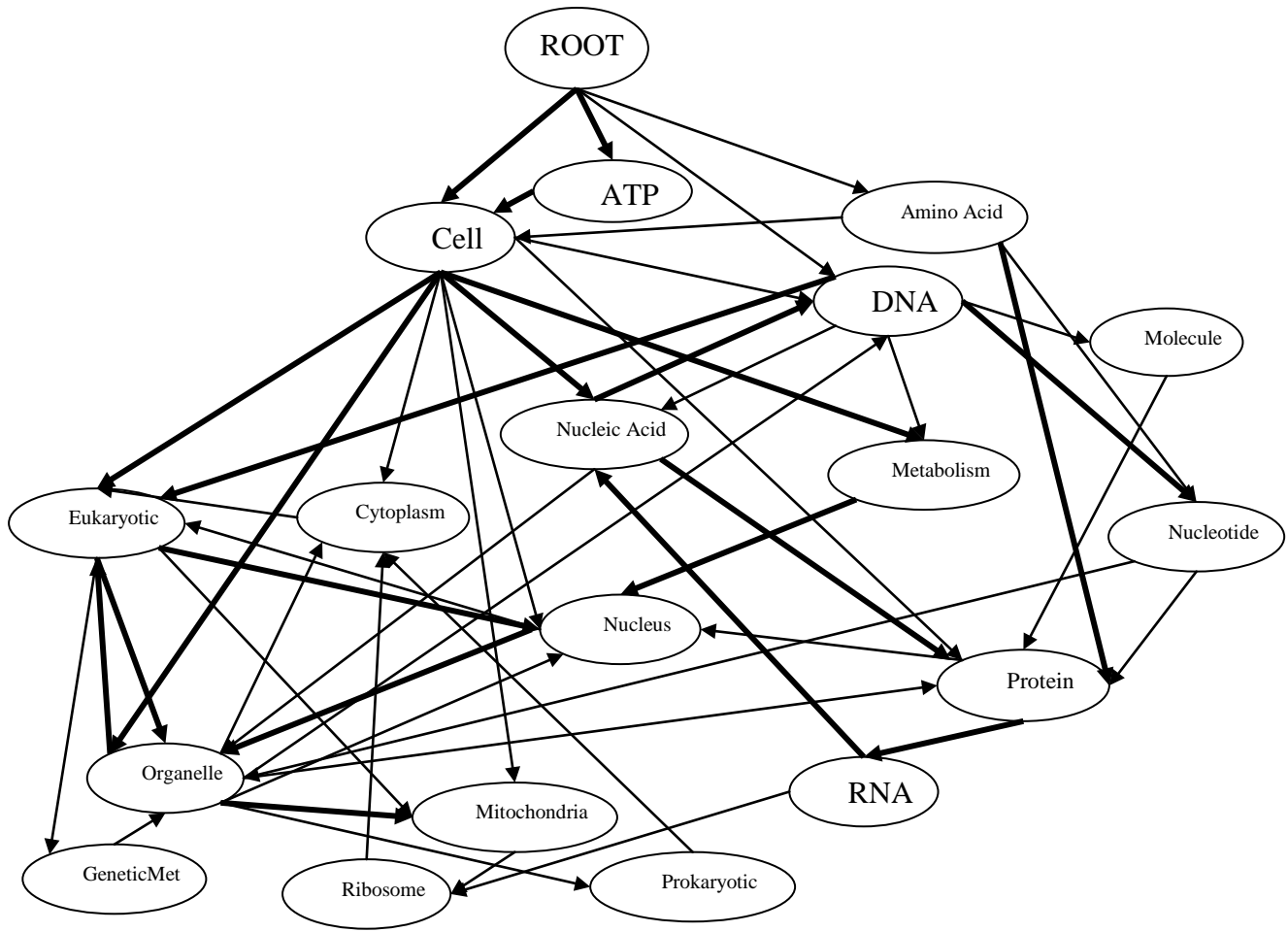

**Fig 6.1: Frequent Pattern Graph**

Next we will try to find association between the nodes or the terms by applying the knowledge obtained from the question answer session of the contact mode classes. We will create transaction from each question answer and map the frequency of these transactions into the branches to form the associations. Let $S= \{s_1, s_2 \ldots s_k\}$ is the set of all terms used by the teacher while delivering the entire lecture on a particular topic. The 'what is' type questions are commonly answered by the term definitions, so in our approach we have taken only such answers to create transactions. $T= \{t_1, t_2 \ldots t_m\}$ is the set of all successful answer made by the teacher on $Q= \{q_1, q_2 \ldots q_m\}$ which is set of questions made by the students and $t_i$ and $q_i$ has 1 to 1 correspondence. We consider all $t_i$ as set of keywords taken from S. So $t_i \subseteq S$. Any term is said to be independent if it does not have other keyword term, belonging to S, is used to define it. Independent keywords are generally easily understandable by its own definition so they have the higher probability of staying nearer to the root. For example say question $t_1$ is satisfactorily answered with the transaction $\{t_2, t_3, t_4\}$ then we search for a branch in the FP Graph with the sequence $\{t_2 \rightarrow t_3 \rightarrow t_4 \rightarrow t_1\}$. The frequency of $t_i \rightarrow t_{i+1}$ will be increased by 1 if it is a existing edge in graph. If the entire transaction does not match to any branch the we look for match with all possible sub transaction with $t_1$ like here we can create $\{t_3 \rightarrow t_4 \rightarrow t_1, t_4 \rightarrow t_1\}$. Now we will keep on updating the FP Graph based on the records stored from different classes taken to different set of students on that particular topic. For every round we will calculate the frequency value of each edges so that if frequency of any link increases to such a level that the frequency reaches or crosses σ, a predefined threshold value, then convert sequence (→) into association (➔). Any keyword which is amalgamated in nature [set of words] always preceded by its subset hence there is always a trivial association from subset to superset. In our example we have converted only some of the links into association based on our assumptions, shown in fig 6.1, but the actual implementation of this model should have a very large number of classroom data of question answer session recorded and then analyzed to form the transactions. Then if the repeated mapping of these transactions increased the frequency of any link above the threshold value then only that link is converted into an association.

### 6. Artificial Intelligence:

The use of swarm intelligence techniques in e-learning scenarios provides a way to combine simple interactions of individual students to solve a more complex problem. Genetic algorithm and Ant Colony Optimization are two commonly used approaches. In our proposal we have suggested Ant Colony Optimization-based inductive planning for recommending learning paths as it is a heuristic method that has been successfully used to deal with this kind of problems. The ants construct a solution by starting from a given node which is a unknown term. Then, at each construction step it moves along the edges of the graph. Each ant checks the data list in subsequent steps to understand which node is used to define the target term and it chooses among the edges that do not lead to vertices that it has already visited. An ant has constructed a solution once it has reached a known word or the root. At each construction step, an ant probabilistically chooses the edge to follow among those that lead to yet unvisited vertices. The probabilistic rule is biased by pheromone values and heuristic information: the higher the pheromone and the heuristic value associated to an edge, the higher the probability of an ant to choose that particular edge. Once all the ants have completed their tour, the pheromone on the edges is updated. Each edge then receives an amount of additional pheromone proportional to the quality of the solutions to which it belongs.

### 7.1 ACO adaptation:

Each ant starts from a query node (unknown term) $q_k$ and iteratively adds nodes till unvisited node is available in its partial tour. The solution construction terminates once a path from $q_k$ to Root node or any node with known keyword is found. Optimal path should contain maximum number of association in it. If the feasible neighborhood $N_i^k$ does not contain any association, then it chooses the edge which has maximum frequency value and data list of which contains the target term. Each term should come at most once in a whole searching process. This constraint is enforced by the rule that an ant at each construction step chooses the next node only among those it has not visited yet. Each ant $a_k$ contain a memory $M^k$ which contain the nodes already visited, in the order they are visited. This memory is used to define the feasible neighbourhood $N_i^k$ in node i, comprises all nodes that are unvisited. This feasible neighbourhood concept we have implemented via introduction of the data lists with each term, by which one ant can understand whether a particular edge takes part in the solution path or not. The pheromone trail value $\tau\tau_{ij}$ indicate how proficient it has been in the earlier iterations to make a particular move from node $s_i$ to node $s_j$ ( $s_i$ → $s_j$ ). In this paper the pheromone represents the occurrence of association of a term in describing other terms. The trail value is therefore a posterior indication of the desirability of that move. Trail value of a particular move will increase if the number of edges which form the association $a_{ij}$, is increased in tour $T^k$ where k = 1,2…m, k is identification number of ant, and m is total number of ant which contain the move $s_i$ → $s_j$. The attractiveness $\eta_{ij}$ is the heuristic information of the move $s_i$ → $s_j$. Trail value $\eta_{ij}$ is proportional to frequency value $c_{ij}$ of the edge $e_{ij}$.

In an Ant System (AS), m (artificial) ants concurrently build path from query node $q_k$ to the root node or any node with known keyword is found. Initially, ants are put on $q_k$. At each construction step, ant $a_k$ applies a probabilistic action choice rule, called random probability rule, to decide which node to visit next. [7] In particular, the probability with which ant $a_k$, currently at node $s_i$, chooses to go to node $s_j$ is

$$P_{ij}^k = \begin{cases} \dfrac{[\tau_{ij}]^\alpha [\eta_{ij}]^\beta}{\sum_{l \in N_i^k} [\tau_{il}]^\alpha [\eta_{il}]^\beta} & if\ s_j \epsilon\ N_i^k \\ 0 & Otherwise \end{cases}$$

where α and β are two parameters which determine the relative influence of the pheromone trail and attractiveness. In our example we have set both of them to 1 for the sake of simplicity.

After each iteration $t$, i.e., when all ants have completed a solution, trail values are updated by:

$$\tau_{ij}(t) = \rho\tau_{ij}(t-1) + \Delta\tau_{ij}$$

Where $0 < \rho \leq 1$ is the pheromone evaporation rate. The parameter $\rho$ is not used in our approach as the edge frequency once achieved will never be reduced. $\Delta\tau_{ij}$ represents the sum of contributions of all ants that used move $s_i \rightarrow s_j$ to construct their solution. All ants deposit pheromone on the edge they have used in their solution path.

$$\Delta\tau_{ij} = \sum_{k=1}^{m} \Delta\tau_{ij}^k$$

Where m is number of ants and $\Delta\tau_{ij}$ is amount of trail deposited on $e_{ij}$ by ant $a_k$, which can be computed as

$$\Delta\tau_{ij}^k = \begin{cases} Q * C^k & if\ s_{ij}\ \in T^k \\ 0 & otherwise \end{cases}$$

Where $C^k$ is represented by the number of association presents at $T^k$. $Q$ being a constant parameter which we have set to 1.

A Learning path is constructed by applying the constructive procedure to each ant: (1) at the start each ant positioned at the query node $q_k$; (2) use pheromone value and heuristic value to probabilistically construct a path by iteratively adding the node that the ant has not visited yet and has the target term in its datalist ,until a path from $q_k$ to Root node; and (3) if a term $q_u$ in the solution is not well understood by the learner and requires extra clarification then the same algorithm will be applied with target term $q_u$ in place of $q_k$ .It is then repeatedly applied until a termination criterion is satisfied.

### 7.2 Searching Algorithm:

The performance of the algorithm depends on the correct tuning of the parameters, and the relative importance of trail and attractiveness. The algorithm mainly works on two procedures *1) LEARNING_PATH 2) UPDATE_TRAIL*

Initialize: $\eta_{ij} = c_{ij}$ and $\tau_{ij} = \tau_{ij}(0) \quad \forall\ e_{ij}\ \in E$,
    Target_Term = Query_Term;

```
Procedure LEARNING_PATH (Node Target_Term)
{
    FOR each ant k       DO
    Repeat
        WHILE Not (ROOT is Reached) DO
        {
            Choose Next_Term using Probability
                    Distribution P_{ij}^k ;
            Target_Term = Next_Term;
            T^k := T^k + Next_Term;
            Update M^k & N_i^k;
            IF (Next_Term == Association)
                C^k = C^k + 1;
        }
        END FOR
        T := min(T^k)   Where k = 1,2,....m
        UPDATE_TRAIL ();
        IF NOT(END TEST)
        LEARNING_PATH (Node Target_Term);
    Else
            Return T;
}

Procedure UPDATE_TRAIL ()
    {
    FOR each edge e_{ij}  DO
    Repeat
        FOR each ant k       DO
        Repeat
            IF(e_{ij} \in T^k)
                Δτ_{ij}^k = Q * C ;
            Else
                Δτ_{ij}^k = 0 ;
        Δτ_{ij} = Δτ_{ij} + Δτ_{ij}^k ;
    // Update the TRAIL MATRIX
        END FOR
    END FOR
        }
```

## 8. Case study :

Based on our above example we tried to find out a recommended learning path for two terms: 'Mitochondria' and ' Eukaryotic' . The last updated data list that we have used is given below .

| Term | Data List |
|---|---|
| ATP | Mitochondria |
| Amino Acid | Ribosome, Protein |
| cell | Mitochondria, Eukaryotic, Organelle, DNA, Nucleic acid, Cytoplasm, Ribosome |
| Cytoplasm | Mitochondria |
| DNA | Nucleic acid, Eukaryotic, Nucleus |
| Organelle | Mitochondria, Eukaryotic, DNA, Cytoplasm, Nucleus, Ribosome |
| Eukaryotic | Mitochondria, Cytoplasm, Nucleus |
| Metabolism | Eukaryotic |
| Mitochondria | Cytoplasm |
| Molecule | Nucleic acid |
| Nucleus | Eukaryotic, DNA, Cytoplasm |
| Nucleic acid | DNA |
| Prokaryotic | Cytoplasm |
| Ribosome | Cytoplasm |
| Nucleotide | Nucleic acid, Ribosome |
| Protein | Nucleic acid, Nucleus, Ribosome |
| RNA | Nucleic acid, Ribosome |
| GeneticMet | Nucleus |

**Table 8.1 : Data list of the terms**

**Results:**
I. For *Mitochondria*: the recommendation is {ATP, Cell, Eukaryotic, Organelle }
II. For *Eukaryotic:* the recommendation is {Cell, Metabolism, Nucleus, Organelle}

## 7. Conclusion :

In this paper we have tried to propose an artificial intelligent based approach to find a learning path which may be helpful for a learner to understand the prerequisite terms, after going through which the unknown target term is more easily understandable. The flexibility of the algorithm is that the solution can be changed at learner's choice of keyword or term which is not known, so the solution is personalized. In this paper we have identified the problem of a self learner to learn a term in a e-Learning environment. Then our approach towards solution is three folded; first we have proposed a frequent graph model which can be created from repository of e-learning contents and then we create transaction from real world classroom session of questions answers and updated the frequencies to identify the associations. Finally we have applied the ant colony optimization technique to find a learning path between two nodes one unknown and the other is either known or comprehensive. The possibility of wrong path selection which is generally biased by the presence of significant number of association in the branches, is nullified by the introduction of the data list in each node.

**Souvik Sengupta** is an Assistant Professor in the Computer 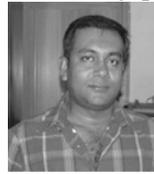 Science Department of Bengal Institute of Technology, India . He has obtained his Master of Technology form West Bengal University of Technology, India in 2008. He is currently performing his research work under the PhD program of the Department of Information Technology of the university of Calcutta ,India. He has 9 international jounal and conference papers published in last 3 years. He has also served  as a visiting faculty in the MTech program of Department of  Computer Science of National Institute of Technical Teachers' Training and Research, Kolkata, India.  His areas of research interests include Web–based Learning, Multimedia Database, Data mining,  Software engineering, Artificial Intelligence etc.

**Sandipan Sahu** is an Assistant Professor in the Computer 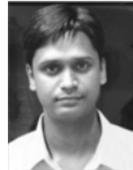 Science Department of Bengal Institute of Technology, India . He has obtained his Master of Technology form West Bengal University of Technology, India in 2007. He has 2 international and national conference paper published. His areas of research interests include Web–based Learning, Data mining, Artificial Intelligence, Image Processing etc.

**Ranjan Dasgupta** is Professor & Head, Dept. of CSE. 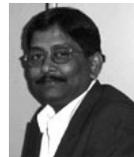 National Institute of Technical Teachers' Training & Research, Kolkata, India, Previously he was with Jadavpur University, and also served several Indian IT companies, for five years. He received his B. Tech from,Dept. of Radio Physics & Electronics, M. Tech, & Ph.D from Dept. of Computer Science, University of Calcutta,India. His areas of research interests include software,engineering, databases, GIS, distributed computing. Dr.Dasgupta has published around 30 research papers in different International Conferences & Journals. He has also served several other Universities & national & international bodies, World Bank Assisted Projects in various capacities.